# Noise figure and photon probability distribution in Coherent Anti-Stokes Raman Scattering (CARS)

**D. Dimitropoulos, D. R. Solli, R. Claps [1], and B. Jalali**
*Department of Electrical Engineering*
*University of California, Los Angeles CA90095-1594*
*jalali@ee.ucla.edu*

[1] *Present address : Neptec Optical Solutions, Fremont CA 94539*

**Abstract:** The noise figure and photon probability distribution are calculated for coherent anti-Stokes Raman scattering (CARS) where an anti-Stokes signal is converted to Stokes. We find that the minimum noise figure is ~ 3dB.

## 1. Introduction

Coherent anti-Stokes Raman Scattering (CARS) is a nonlinear, parametric process with a wide application in spectroscopy and in data conversion [1,2,3]. The problem that we examine is the following: how are the signal-to-noise ratio (SNR) and photon statistics of an input signal at one sideband (say the Stokes frequency) modified when the signal is converted to the other sideband (say the anti-Stokes frequency)?

The CARS process necessarily introduces extra noise in the converted signal because it occurs through the Raman nonlinear optical susceptibility. Since the process involves the interaction of damped optical phonons with the incident frequency in a medium, noise is introduced in the system. Therefore, in order to formulate the process mathematically, one needs a Langevin noise source to correctly model the dissipation. In what follows we use this principle to calculate some of the noise properties of the converted optical signal. We calculate the noise figure and for the first time we obtain an analytic expression for the photon statistics.

## 2. CARS quantum mechanical equations with noise

The equations describing the evolution of the operators for the Stokes, anti-Stokes and optical phonon waves are:

$$\frac{d}{dx}\hat{a}_2 = -j\xi a_1 \hat{q}^+ \quad , \quad \frac{d}{dx}\hat{a}_3^+ = jr\xi a^*_1 \hat{q}^+ \exp(-j\Delta\beta \cdot x) \tag{1a,b}$$

$$\frac{d}{dt}\hat{q}^+ = j\xi a^*_1 \hat{a}_2 + jr\xi a_1 \hat{a}_3^+ \exp(j\Delta\beta \cdot x) - \gamma\hat{q}^+ \tag{1c}$$

where $\hat{a}_2$, $\hat{a}_3$, $\hat{q}$ are the creation operators for the Stokes, anti-Stokes and phonon fields with commutators:

$$[\hat{a}_2(f), \hat{a}_2^+(f')] = \delta(f-f'), \quad [\hat{a}_3(f), \hat{a}_3^+(f')] = \delta(f-f'), \quad [\hat{q}(x), \hat{q}^+(x')] = \delta(x-x'). \tag{2}$$

The incident (pump) optical wave amplitude is $a_1$ (c-number), $\gamma$ is the phonon damping rate, $r = \sqrt{f_3/f_2}$, and $\Delta\beta = 2\beta_1 - \beta_2 - \beta_3$ is the wavevector mismatch between the waves. Apart from the damping term in the optical phonon equation all other equations are derivable from a Hamiltonian (Kartner et al [4] presents the form of the Hamiltonian). Due to the damping term, the commutator of the phonon mode will not be invariant in time unless a noise source is introduced. For the damping term in equation (1c), the noise source operator that must be added is $\sqrt{2\gamma}\hat{N}$ where $[\hat{N}(f), \hat{N}^+(f')] = \delta(f-f')\delta(x-x')$. The justification and derivation of the result can be found for example in [5]. The noise operator acts on states of the "noise reservoir'.

## 3. Noise figure for signal conversion at perfect phase-matching and photon number distribution

We calculate the noise figure and photon number distribution at the peak of the resonance, where the Stokes gain is maximized for $\Delta\beta = 0$. The solution for the Stokes wave is:

$$\hat{a}_2(x) = \frac{r^2 - \exp(-\beta x)}{r^2 - 1}\hat{a}_2(0) + r\frac{1 - \exp(-\beta x)}{r^2 - 1}\hat{a}_3^+(0) - j\sqrt{g}\int_0^x dx' \exp(-\beta(x-x'))\hat{N}^+(x'), \quad (3)$$

where $\beta = g(r^2-1)/2$, and $g = 2(\xi^2/\gamma)|a_1|^2$. To calculate the noise figure (SNR relative to the shot noise of the input) we need to calculate the mean photon number and the photon number fluctuations as a function of x. The noise reservoir is taken to be in the ground state and we assume the anti-Stokes input is a coherent state. This situation corresponds to the state state $|0\rangle_2|a\rangle_3|0\rangle_R$ where $|\,\rangle_2$, $|\,\rangle_3$, and $|\,\rangle_R$ denote the Stokes, anti-Stokes, and "noise reservoir" states.

When $gx \gg 1$ and in the opposite limits $\beta x \ll 1$ or $\beta x \gg 1$ we find the noise figure:

$$F_{\min} \cong 2. \quad (4)$$

The results deviate from the ideal optical amplifier noise figure by a factor which depends on the ratio of the photon energies of the Stokes and anti-Stokes waves. Therefore, the extra deterioration (or improvement in the case of Stokes to Anti-Stokes conversion) in the *SNR* is due to the change in photon energy at the conversion process.

We next show how to calculate the photon number distribution function for the general case in which $\hat{a}_2(x) = A\hat{a}_2(0) + B\hat{a}_3^+(0) + N\hat{N}^+$ (assume $[\hat{a}_2(0),\hat{a}_2^+(0)] = [\hat{a}_3(0),\hat{a}_3^+(0)] = [\hat{N},\hat{N}^+] = 1$, the substitutions that need to be made after are straightforward). We first calculate the characteristic function of the probability distribution :

$$\langle \exp(-jk\hat{a}_2^+(x)\hat{a}_2(x))\rangle = {}_R\langle 0|_3\langle a|_2\langle 0|\exp(-jk\hat{a}_2^+(x)\hat{a}_2(x))|0\rangle_2|a\rangle_3|0\rangle_R \quad (5)$$

After some manipulation, we obtain the result:

$$\langle \exp(-jk\hat{a}_2^+\hat{a}_2)\rangle = e^{-x'}\frac{e^{jk}}{1+y(e^{jk}-1)}\exp\left(\frac{x'}{1+y(e^{jk}-1)}\right), \quad (6)$$

where $y = |A|^2$, and $x' = |Ba|^2/y$. The characteristic function is a periodic function of $k$ with a period $2\pi$ and can be represented with a Fourier series. The coefficients of the series give the probability distribution of the photon number states. The probability that the photon number $N = \hat{a}_2^+(x)\hat{a}_2(x)$ has a value $n$ given that the anti-Stokes input is a coherent state with amplitude $a$ is :

$$P(N=n|a) = (1/2\pi)\int_0^{2\pi} dk\, e^{jkn}\langle \exp(-jk\hat{a}^+\hat{a})\rangle. \quad (7)$$

We can change variables $z = e^{jk}$ and perform the complex integral to find:

$$P(N=n|a) = e^{-x'}\sum_{m=0}^n \frac{(x'/y)^m}{ym!m!}\frac{n!}{(n-m)!}(1-y^{-1})^{n-m}. \quad (8)$$

With the knowledge of the distribution function one can now calculate the information capacity of the CARS process.

## 4. Conclusions

We have calculated the SNR degradation and photon statistics for signal conversion through the CARS process. We find that with perfect phase matching the best noise figure is $F \sim 3dB$ since the conversion process introduces noise to the signal. The photon probability distribution of the converted signal was also calculated when the input is a coherent state.